# Signatures of non-Markovian turbulent transport in Reversed Field Pinch plasmas

F. Sattin[#], R. Paccagnella[§ &], F. D'Angelo[$]

Consorzio RFX, Associazione Euratom-ENEA sulla Fusione

Corso Stati Uniti 4, 35127 Padova, ITALY

[§] Consiglio Nazionale delle Ricerche, Italy

**Abstract**

Transport of field lines is studied for a realistic model of magnetic field configuration in a Reversed Field Pinch. It is shown that transport is anomalous, i.e., it cannot be described within the standard diffusive paradigm. To fit numerical results we present a transport model based upon the Continuous Time Random Walk formalism. Fairly good quantitative agreement appears for exponential memory functions.

[#] E-mail address: fabio.sattin@igi.cnr.it
[&] E-mail address: roberto.paccagnella@igi.cnr.it
[$] E-mail address: ferdinando.dangelo@igi.cnr.it

# 1. Introduction

It has been known since long that the various instabilities arising in a magnetically confined plasma may deeply alter the transport of matter and energy, making it to deviate from the standard collisional diffusion paradigm. A thorough understanding needs taking into account the full set of interactions between the plasma particles, the self-consistent magnetic field and electric fields: a formidable task. Hence, researchers often tackle a reduced version of the problem, where the particles act as passive tracers advected by background **E**, **B** fields whose behaviour is known through independent models. Even these simplified pictures can be highly sophisticated and contain a lot of physics, to the extent that the interpretation of raw output of calculations is far from trivial. In the past years, people have indeed become growingly convinced that, in order to correctly understand the evidence coming from these numerical experiments, one had to go beyond the standard paradigm of transport as a statistical Markovian (memoryless) process: The ansatz of a Markovian process translates into phenomenological transport equations that are of the diffusion-convection type. Hence, the only macroscopic effect of turbulence, within the Markovian framework, seems that of giving numerical estimates for the diffusion and convection coefficients that differ from the purely collisional ones; while, going over the Markovian paradigm affects the form of the equations themselves.

We make quantitative the previous statement by briefly recalling Balescu's treatment of the subject [1]. Let us suppose of describing the time- and space-evolution of the density of some quantity, $f(x,t)$, evolving within a background field that, because of its complexity can be described only on a statistical (stochastic) fashion. In [1], this process is described using the Hybrid Kinetic Equation

$$\frac{\partial}{\partial t} f(x,t) + \mathbf{v}(t) \cdot \nabla_x f(x,t) = 0 \qquad (1)$$

where **v** represents a realization of the stochastic field. A key passage, then, is to divide the density $f$ into an average and a fluctuating part: $f(x,t) \equiv n(x,t) + \delta f(x,t)$, and replace this expression into Eq. (1), which may then be split into two coupled equations for $n$ and $\delta f$. One is usually not so much interested in the wildly fluctuating part $\delta f$, rather to the mildly varying $n(x,t)$. Balescu showed that it is possible to get rid of the explicit presence of $\delta f$ into the equation for $n(x,t)$ and write the evolution equation for this latter quantity in the form

$$\frac{\partial n}{\partial t}(x,t) = \int_0^t \Lambda(t-\varsigma)\frac{\partial^2 n}{\partial x^2}(x,\varsigma)d\varsigma \qquad (2)$$

where the kernel $\Lambda$ is made by autocorrelations of the stochastic variable **v**. This expression holds under the rather general assumption of systems slowly varying in space and symmetrical with respect to reflection $x \rightarrow -x$. In the most general case, the rhs of Eq. (2) becomes an integral equation over both temporal and spatial coordinates. Eq. (2) is an integro-differential equation, which made its entrance in the statistical physics within the theory of Continuous Time Random Walks (CTRW), independently and prior to the present treatment (see chapter 12 in [1] and references therein). Depending on the form of the kernel $\Lambda$, Eq. (2) may or may not represent a Markovian process. In the limit of vanishingly small autocorrelation time for **v**, $\Lambda$ approaches a Dirac delta: $\Lambda(t) \rightarrow \delta(t)$, and the Markovian limit, yielding the standard diffusive process, is recovered. In the general case, the behaviour of density $n$ at time $t$ depends upon its previous behaviour over a finite time interval.

The analytical form for $\Lambda$ depends upon the properties of the background field. However, the connection between the two quantities is fairly involved and one can hardly hope to explicitly compute $\Lambda$ from first principles. Rather, the other way round is usually followed: one builds a fixed (possibly time-dependent) background on top of which a preassigned distribution of test particles is let to evolve. The time behaviour of various moments of the distribution yields the transport properties: of particular importance is the second moment of the spatial displacement: $<\Delta x^2> \propto t^\mu$, with the exponent $\mu$ which differs from unity for anomalous transport, i.e., within the picture (2), when the kernel $\Lambda$ has a finite memory. The power-law dependence from time, $\Lambda \approx t^{-a}$, is extremely relevant from this point of view, and hence most often investigated, since it may yield a non-diffusive transport even asymptotically for $t \rightarrow \infty$. Some examples of this kind of studies are [2-14]. However, in principle other forms for $\Lambda$ are possible. Inferring a plausible form for $\Lambda$ from the computed moments will be one of the purposes of the present paper.

Tokamaks are, to date, the most successful magnetic fusion devices [15,16]. The plasma is kept confined in their interior by strong magnetic fields, induced by external coils: plasma particles are sticked to magnetic field lines by the Lorentz force. In order to avoid diffusion of the plasma towards the outside, it is needed that the field lineslie on nested and closed surfaces, and the only suitable configuration takes the shape of a torus. In order to be stable, this configuration needs a magnetic field with both a poloidal and a toroidal component. The equilibrium magnetic field is therefore two-dimensional. Tokamaks have in their core high temperatures and pressures, hence a large amount of free energy available to drive instabilities of all sorts. Main instabilities are the pressure-gradient-driven modes that generate space-and-time-fluctuating electric fields. The resulting stochastic velocity appearing in Eqns. (1,2) comes from this fluctating electric field and is the drift velocity $\mathbf{v}_E = \mathbf{E} \times \mathbf{B}/B^2$. The level of magnetic turbulence is, instead, relatively low. Notice, however, that the presence of magnetic fluctuations makes three-dimensional the overall geometry of the magnetic field. While a two-dimensional configuration insures the existence of closed magnetic surfaces, adding a radial component breaks a fraction of them. KAM theorem, however, assures that most of the surfaces are retained as long as the perturbation is sufficiently small.

Reversed Field Pinches (RFPs) represent an alternative line of magnetic confinement devices [16]. They are, like tokamaks, toroidal devices. Unlike tokamaks, the final magnetic configuration is only partially induced from the outside, and mostly self-generated by the plasma itself through the generation of flowing currents. A peculiar feature of this device (after which its name comes) is that the toroidal component of the magnetic field reverses its sign close to the edge. Up to now, performances of existing RFPs have been plagued by a high level of magnetic turbulence (about two orders of magnitude more than tokamaks'), that degrades the quality of confinement: average temperatures in RFPs are 10- times lower than in tokamaks, for comparable plasma densities. Plasma pressure is correspondingly lower, hence pressure-diven turbulence is not the main actor here.

In the RFP the three-dimensional magnetic perturbations can break a relevant fraction of the magnetic surfaces, hence allowing for chaotic motion of field lines throughout most of the plasma volume.

As long as one neglects drifts and collisions, particle and field line motion may be identified. The standard reference for this subject is the paper by Rechester and Rosembluth (RR) [17], where they derived an analytical expression for magnetic field line diffusivity in the case of homogeneous and fully developed turbulence (A more extensive treatment of particle transport in stochastic magnetic fields can be found, e.g., in the paper by Krommes, Oberman and Kleva [18]). However, magnetic profiles in real devices do not share all of the features of the RR model: first of all, their topology is fairly more complicated, mixing fully chaotic regions with others where at least a partially regular structure does exist. Wherever Lyapunov exponents are small, we are led back to Eq. (2), i.e., the Markovianization implicitly imposed in RR does not hold. In second place, due to the obvious fact that any physical device is finite, the hypothesis of a perfectly homogeneous background breaks down, and finite-size effects do modify field line wandering.

A few years ago, D'Angelo and Paccagnella [4,5] (hereafter referred to as DP, DP1) devoted a study, both analytical and numerical, to assess the transport properties of field lines in a quite realistic model of magnetic field for a RFP. It allowed giving a nice account of the long-time properties of the transport, when finite-size effects begin to dominate. The mathematical apparatus of CTRW's was not so widely diffused in theoretical plasma physics at that time, hence short-time behaviour was interpreted just in terms of the RR model, i.e. as a diffusive process. The agreement between numerical results and RR's theoretical predictions was not perfect, hence it was clear that signatures of "anomalous transport" were present in the simulations. It is the purpose of the present paper to revisit D'Angelo and Paccagnella's study in the light of the new knowledge gained in these years, trying to unify under a single formalism the transport description in different regimes i.e. from Markovian to non-Markovian. Next section gives a brief summary of the numerical model for the zeroth order magnetic field, the perturbations, and the field line evolution equations. In section 3 we present a few results, pointing to the features that make them appear as consequences of an intrinsically non-Markovian process. Conclusions are drawn in Section 4.

**2. Field line Hamiltonian**

A standard approximation is to approximate the full toroidal shape of the plasma device with a cylindric one, with periodic boundary conditions. In cylindrical coordinates $(r, \theta, \phi = z/R)$ the field line equations can be written as [19]:

$$\frac{dr}{dL} = \frac{B_r}{B}, \quad \frac{d\theta}{dL} = \frac{B_\theta}{rB}, \quad \frac{d\phi}{dL} = \frac{1}{R}\frac{B_\phi}{B} \quad , \tag{3}$$

where $L$ is the coordinate along the magnetic field line ("generalized time")[1], $B = |\mathbf{B}|$, $R$ is the torus major radius, and all lengths are normalized to plasma minor radius, $a$. We will use a perturbative approach, so the total magnetic field may be decomposed as $\mathbf{B} = \mathbf{B}_0 + \delta\mathbf{B}$, where

$$\mathbf{B}_0 = B_0 \left( b_\theta(r)\hat{\mathbf{e}}_\theta + b_\phi(r)\hat{\mathbf{e}}_\phi \right) , \quad \delta\mathbf{B} = B_0 \beta(r) \sum_{m,n} b_{mn} \sin(m\theta - n\phi + \mu_{mn}) \hat{\mathbf{e}}_r \quad (4)$$

The three versors $\hat{\mathbf{e}}_\theta, \hat{\mathbf{e}}_\phi, \hat{\mathbf{e}}_r$, are directed respectively along the poloidal, toroidal and radial direction. Hence, the umperturbed magnetic field $\mathbf{B}_0$ has only poloidal and toroidal components. On the other hand, the small perturbation $\delta\mathbf{B}$ is just along the radial direction (components of $\delta\mathbf{B}$ along the two other directions have negligible influence along transport, since amount to adding a small increment to the much larger contribution provided by $\mathbf{B}_0$).

The dimensional constant $B_0$ carries over the dimensionality of the field. The perturbation $\delta\mathbf{B}$ has a peculiar form: it is built from a linear combinations of modes, with typical amplitude $b_{mn}$ and a radial profile given by the function $\beta(r)$, the same for all modes: $\beta(r) = r(1-r)$ The advantage of this expression is that can easily be cast into Hamiltonian form. The Hamiltonian can then be written as:

$$H = H_0(I) + \varepsilon\, H_1(I, \varphi, \theta) = \int^I \omega_O(I')\, dI' + \varepsilon\, f(I) \sum_{mn} \frac{b_{mn}}{n} \cos(m\theta - n\phi + \mu_{mn})$$

(5)

in the action-angle representation, where the action $I = r^2$ is physically just the square of the radial coordinate, while $\phi$, the angle, corresponds to the toroidal angle. The poloidal angle $\theta$ instead can be seen as the time-like variable for this system. Therefore the Hamiltonian equations of motion are:

$$\frac{dI}{d\theta} = -\varepsilon\, f(I) \sum_{mn} b_{mn} \sin(m\theta - n\phi + \mu_{mn}),$$

$$\frac{d\phi}{d\theta} = \omega_0(I) + \varepsilon\, f'(I) \sum_{mn} \frac{b_{mn}}{n} \cos(m\theta - n\phi + \mu_{mn})$$

(6)

---

[1] Throughout this paper we are using two different symbols for time: the physical time, $t$, and the generalized time (actually, a length), $L$. This choice is related to the fact that $t$ is the most logical variable when dealing with transport equations (such as Eqns. 1,2), while $L$ is more natural within the Hamiltonian formulation (3-7). The ambiguity is more apparent than real, infact $t$ and $L$ are linearly related: $L = u \times t$, where $u$ is the velocity with which one moves along the field line. Through a suitable choice of time scale, we may set $u = 1$, hence $t$ and $L$ become numerically equal.

The function *f(I)* is related to β(r) and gives the action dependence of the perturbative Hamiltonian, $\omega_0(I) = q(I)$ is the safety factor, and ε is a scaling factor, giving the order-of-magnitude of the perturbative Hamiltonian. The $\mu_{mn}$'s are phases that can be taken randomly. All these functions must be given in advance to calculations. We use the same expressions as in DP:

$$q(I) = q_0\left(1 - \frac{I}{I_*}\right),$$
$$f(I) = 2I\left(\sqrt{I} - 1\right)$$
(7)

These expressions derive from assuming truncated Bessel profiles for zeroth order magnetic field (taking into account the relation between *I* and the radius): Bessel functions are good first-order approximations for the magnetic field in a RFP [16]

The first equation of (7), if $I_* < 1$, insures a monotonic decreasing behaviour of *q* from $q = q_0 > 0$ at the centre to a negative value at the edge, consistently with RFP profiles.

We remark (especially for readers outside fusion research) that the particular shape of the *q* profile is at the origin of the possible chaotic behaviour of the magnetic lines wandering in the RFP device. In fact, the simultaneous presence of many resonances with *n>m* (since $q_0 << 1$) in the plasma can easily lead to the violation of a Chirikov criterion, where the Chirikov parameter is calculated as the ratio of the magnetic island size to the radial separations of neighbouring resonances.

The shape of *f*, instead, guarantees that the perturbations have vanishingly small amplitudes at the centre and at the edge. The mode spectrum is chosen to vary as: $b_{mn} = (n_{min}/n)^2$, where $n_{min}$ is the smallest of the *n* numbers used in the simulations (we used *m* = 1, and *n* = 7,...,30. This roughly agrees with the spectrum of modes found in present-day Reversed Field Pinches [20]). Equations (6) are solved together with the second of (3), to give *I(L)*, φ*(L)*.

The model given by Eqns. (3-7) represents just a rough approximation for a realistic model of the magnetic field for a RFP. It is perhaps interesting to stress that, recently, a much more sophisticated model has been developed [21,22]. That model predicts profiles of zeroth order magnetic fields as well as of perturbations, by solving the MHD force balance equation in the presence of saturated resonant instabilities, which is the standard condition in a fully developed RFP. Also, the absolute value of all of the fields can be matched to experimental values by using edge magnetic field measurements as boundary conditions. Some sample of the profiles for magnetic field perturbations can be found in the papers [21,22].

However, we do not need all the accuracy provided by this more sophisticated model. Instead, Eqns. (3-7) are easier to manipulate and provide us with few but effective handles to make parametric studies. In particular, we can start from a perfectly integrable situation for $\varepsilon = 0$ and move to a more and more chaotic one by increasing it. Furthermore, the intertwining between this parameter, the scaling form of $b_{mn}$'s, and the *q*-profile allows having regions with different levels of stochasticity: one might move with continuity from a region where Chirikov overlapping criterion for the onset of stochasticity is satisfied, to another where it is only marginally, or even is not satisfied at all (for an example, see Fig. 2 in DP).

## 3. Numerical results and their interpretation

*3.1 General case*

The numerical computations follow standard guidelines: Eqns. (6) together with (3) are numerically solved for a set of initial conditions (fixed *I*, random $\phi$), the mean quadratic dispersion is evaluated as a function of a generalized time. In all simulations we used $I_{start} = 0.5$, i.e., roughly at the middle of the radius. From the previous study DP1 we did not expect qualitative differences when varying this parameter, except when we move fairly close to the centre ($I = 0$) or the edge ($I = 1$). But, then, finite-size effects should arise since the very start of the simulations, making them not interesting. The amplitude of the perturbations, $\varepsilon$, appeared to be a more interesting parameter, hence we varied it slightly around $10^{-2}$, which is the typical order-of-magnitude of normalized RFP magnetic field fluctuations.

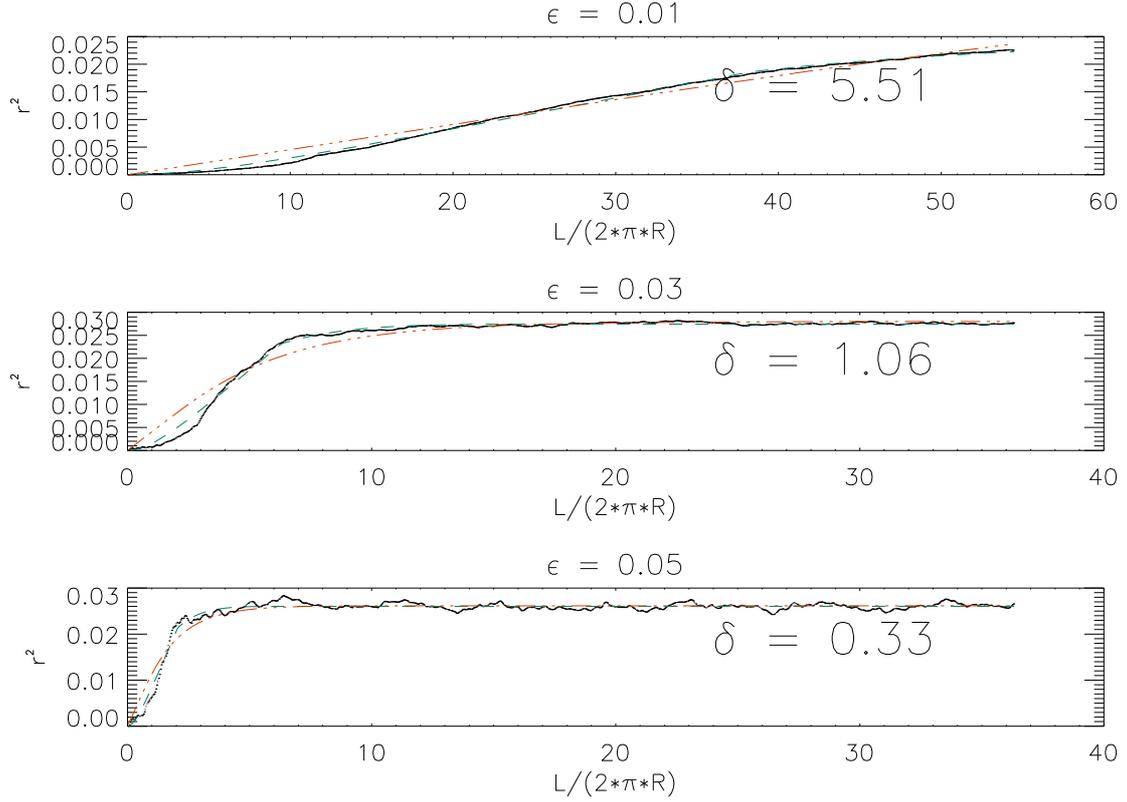

Figure 1. Mean square displacement versus generalized time. Parameters are $\varepsilon = 10^{-2} \div 5\times10^{-2}$ (from top to bottom), $m =1$, $n = 7,...,30$, $I_* = 0.64$, $q_0 = 0.17$. The average is performed over 5000 trajectories. The abscissas are the field line length $L$ normalized to the torus circumference $2\pi R$. The broken thick line is the numerical result; the dashed curve the fit from Eqns. (11,15,16) of the present work; the chain curve the fit from Eq. (23) of the paper DP (Markovian approximation). For each plot the value of the characteristic time scale (the parameter $\delta$ appearing in Eqns. 14-16) is displayed.

Figure 1 shows some samples of our results. We wish to point out that the purpose of this work was addressing the existence of some peculiar physical effects, not carrying on a full-fledged parametric study.

Let us point to the most obvious features: first of all, there is the saturation of mean displacement at long times, when field lines have filled the whole available region, i.e. when a field line, started at a given initial radius, reach the material wall containing the plasma (at $r = b$) Eventually, due to the structure of the chaotic region, the line can also be trapped in a smaller region, $r < b$, but RFPs are built with metallic shells that shield the magnetic fields and prevent field lines to reach larger radii. At small times, it is clearly discernible a trend more than linear, and approximately quadratic. It definitely rules out the possibility for the field line to perform a classical random walk (red chain line) and hints instead to a ballistic motion. Indeed, an intuitive picture may perhaps be provided: even in a chaotic environment, the trajectory of a test particle at a later time $t$ is decorrelated from its previous motion at *t'* only when $t - t' \geq 1/\lambda$, and $\lambda$ is the largest Lyapunov exponent. As long as decorrelation does not occur, time evolution of the position of a particle must be considered as well known: $x = F(t)$, with *F* a well-defined function; and expanding around zero: $x(t) \approx x(0) + F'(0)\,t$, from which the quadratic trend is recovered: $(x - x(0))^2 \propto t^2$. It is already known from previous study (see Fig. 9 in DP) that, for this system, $\lambda$ is of order $10^{-2}$, giving several tens of steps before chaotic separations of trajectories develops. However, over these typical times, the statistical ensemble has already evolved significantly into the system. We are, therefore, in a situation where no clear-cut separation can be made between the microscopic (small-scale) dynamics, and the large-scale one.

Let us now move to interpreting these results in terms of an expression of the kind (2). In this case, *n* is the radial density of field lines. Just like in DP, we neglect the curvature terms, and approximate our system with a slab bounded between $x = -b$ and $x = +b$. Particles are placed initially in the middle of this region, hence the initial condition is $n(x,0) = \delta(x)$. The boundary conditions, on the other hand, are determined by the impossibility of field line of crossing the boundaries, hence the radial flux must be zero there: $\partial n/\partial x |_{x=-b} = \partial n/\partial x |_{x=+b} = 0$. This is related to the fact that we chose radial perturbation to be zero at the boundary of the device. Hence, good magnetic surfaces do exist there. These boundary conditions call for a solution of (2) through separation of the variables:

$$n(x,t) = \sum_{i=0}^{\infty} X_i(x) T_i(t) \tag{8}$$

The spatial functions are promptly determined:

$$X_i(x) = (1 + \delta_{i0}) \frac{1}{2b} \cos\left(\frac{i\pi x}{b}\right) \tag{9}$$

under the condition $T_i(0) = 1 \; \forall i$. Hence, we get an equation for the purely temporal part

$$\frac{dT_i}{dt}(t) = -\left(\frac{i\pi}{b}\right)^2 \int_0^t d\varsigma\, \Lambda(t-\varsigma) T_i(\varsigma) \tag{10}$$

Using Eqns. (8,9) we can formally compute the mean square displacement:

$$\begin{aligned}
<\Delta x^2>(t) &= \int_{-b}^{+b} dx\, x^2 n(x,t) = \sum_{i=0}^{\infty} T_i(t) \int_{-b}^{b} dx\, x^2 X_i(x) \\
&= \frac{b^2}{3} + \sum_{i>0} (-1)^i \left(\frac{2b}{i\pi}\right)^2 T_i(t)
\end{aligned} \tag{11}$$

Equation (11) is just a restatement of Eq. (23) in DP. The difference is embedded in the basis functions $T_i$, which are now unknown. Hence, we cannot naively solve Eq. (10) or (11) for $\Lambda$. Instead, we have to take a longer tour. First of all, we guess a reasonable analytical expression for $\Lambda$ (containing some free parameters). Then, we replace this $\Lambda$ in Eq. (10) and solve for $T_i$. The resulting set of functions is inserted into Eq. (11) and the free parameters adjusted till we get a good match between the empirical expression on the lhs of (11) and the theoretically derived one on the rhs.

Physically, correlations must go to zero for very large time delays, $\Lambda(t) \to 0$, for $t \to \infty$, hence we choose $\Lambda$ as an exponentially decreasing function:

$$\Lambda(t) \approx \frac{K}{\delta} e^{-\frac{t}{\delta}}. \qquad (12)$$

By varying the parameter $\delta$, in principle a wide set of particular cases can be taken into account (practically, all relevant ones). In particular, the Markovian limit may also be recovered (the Dirac delta is the limit of Eq. 12 for $\delta$ going to zero).

The integro-differential equation (10) can be translated in algebraic form through Laplace transform:

$$s\widetilde{T}_i(s) - 1 = -\left(\frac{i\pi}{b}\right)^2 \widetilde{\Lambda}(s)\widetilde{T}_i(s) \qquad (13)$$

Tilded quantities stand for the Laplace transformed ones, and

$$\Lambda(t) = \frac{K}{\delta} e^{-\frac{t}{\delta}} \to \widetilde{\Lambda}(s) = K\frac{1}{1+\delta s}. \qquad (14)$$

Inserting Eq. (14) into Eq. (13), with the position $\lambda_i = i^2\left(\frac{\pi}{b}\right)^2 K$, yields

$$\left(s + \frac{\lambda_i}{1+\delta s}\right)\widetilde{T}_i(s) = 1 \to \widetilde{T}_i(s) = \frac{1+\delta s}{s + \delta s^2 + \lambda_i}$$
$$\to T_i(t) = e^{-\frac{t}{2\delta}}\left[\cosh\left(\omega\frac{t}{2\delta}\right) + \frac{1}{\omega}\sinh\left(\omega\frac{t}{2\delta}\right)\right] \qquad (15)$$

with $\omega = \sqrt{1-4\delta\lambda_i}$. Notice that $\omega$ may be imaginary. In that case, putting $\omega = i\gamma$ into Eq. (15) yields

$$T_i(t) = e^{-\frac{t}{2\delta}}\left[\cos\left(\gamma\frac{t}{2\delta}\right) + \frac{1}{\gamma}\sin\left(\gamma\frac{t}{2\delta}\right)\right] \qquad (16)$$

which has the form of a damped oscillator.

In particular, for small values of the argument, $T_i \approx 1 - \frac{\lambda_i}{2\delta}t^2$, and we get the sought quadratic trend. Checking that Eq. (15) yields the correct result even for $\delta \to 0$ is rather complicated if one uses the final expression on the second line for $T_i$; however, it is straightforward if one puts $\delta = 0$ *before* performing the inverse Laplace transform, and the result is $T_i = \exp(-\lambda_i t)$, just like in DP.

In Fig. (1) we have plotted the fit (11) using the form (15) for the $T$'s. Because of the $i^2$ term appearing in the denominator of the sum in (11), the contribution of higher-$i$'s becomes negligible after the first terms. In actual calculations, we retained terms up to $i = 21$ included. It is apparent how both the small-$t$ as well as the large-$t$ part of the curves are reproduced accurately. This cannot give us guarantee that Eq. (12) is the true analytical form for $\Lambda$, but gives us much confidence in it. In particular, in literature the preferred choice is in favour of a memory function that decays as a power-law. This case, too, leads to analytical solutions for the $T_i$'s in terms of Mittag-Leffler functions [12,23-25]. Our first attempt, hence, was using this kind of memory function. We found that these functions led to several complications, at least in the form currently used and eventually we preferred to discard this choice.

The parameter $\delta$ has the simple meaning of a time scale over which the system reaches saturation. From Fig. (1) we notice that $\delta$ decreases at higher perturbations. This is consistent with the picture of turbulence as a decorrelating mechanism. A faster decorrelation means, in fact, approaching the Markovian limit which, on the basis of Eq. (11), is recovered exactly for $\delta \to 0$. One might wonder whether $\delta$ is somehow related with Lyapunov exponents of the system. Indeed, from the form of the expression where $\delta$ appears, a relation of the kind $\delta \approx 1/\lambda$ might be expected. Hence, $\delta$ is related to the characteristic time over which two trajectories decorrelate completely. We carried on, therefore, the computation of the maximum Lyapunov exponent under the same conditions of Figure (1), Fig. (2) shows $\delta$ plotted versus $\lambda$. It is just obvious how the opposite dependence between $\delta$ and $\lambda$ is recovered, although hints to a stronger power-law trend, than a simple inverse trend. In Fig. (2) we have added also the plot of the Lyapunov exponent versus perturbation amplitude, just to remark the intuitive fact that the former is an increasing function of the disorder.

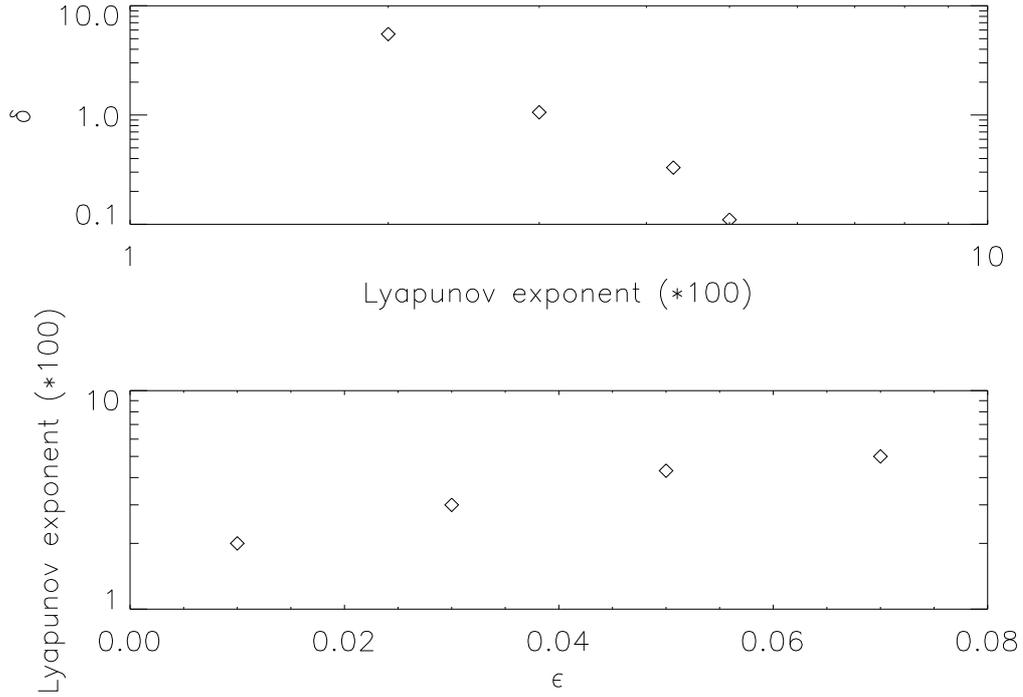

Figure 2. Upper plot, delta parameter appearing in the memory function (12) *versus* Lyapunov exponent, for ε = 0.01,..., 0.07. Lower plot, Lyapunov exponent *versus* ε.

Looking at the underlying different topologies of the field lines in the two cases (see Figure 3), can help in understanding (qualitatively) the different scalings. It is evident that the high ε case corresponds to a well developed chaotic behaviour in the system, while at the lower ε value ordered structures and non-chaotic regions are still present. Therefore the degree of "non locality" of the diffusion process is certainly higher at low ε, where the expression given by Eq.(16) for the $T_i$ basis functions should be used and the simple inverse power law dependence of δ with λ is lost. This observation confirms the fact that a generalized transport theory, such that described in this paper, can really help in studying the transport in different regimes, which correspond to a different level of "non locality" or deviation from a Gaussian statistics.

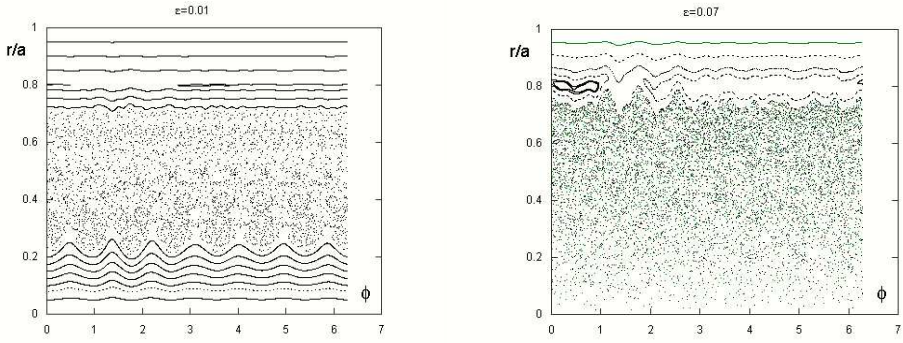

Figure 3. Poincarè plots for the two extremal conditions: left plot, ε = 0.01; right plot, ε = 0.07.

*3.2 Locked modes case*

As a further example of the usefulness of the technique described here, we will focus on the interesting case-for RFP performances-of *locked modes*. This constitutes, from our point of view, another interesting example of deviation from a Gaussian statistics.

In Section 2 we supposed the phases of all modes $\mu_{mn}$ to be random numbers, constant in time and independent between them. However, in almost all RFPs it is found that while the first hypothesis is fairly well fulfilled, the latter is not: modes do preferentially tend to lock so that the relative phases between them are fixed and close to each other [26]. The overall envelope of the perturbation is dubbed *slinky mode:* it features in the toroidal direction a regular pattern of alternating maxima (*i.e.*, locations where the effective perturbation is strongest) and minima (*i.e.* where the destructive interference of the phases leads to a total perturbation that may be even reduced with respect to the single-mode case). We may ask, therefore, what consequences this pattern can have on field line transport. Only small, if any, consequences may arise if we let the line trajectories to start from random locations. This latter randomness, in fact, completely washes out any correlation built in the phases of the modes. Hence, when considering averages over all the available phase space, we can still rely on previous results. However, for particular purposes, we may be interested in investigating the behaviour of a small subset of phase space: in this particular case, trajectories picked up only from within a small interval of toroidal angles. In this case, trajectories are, initially, strongly correlated between them and a collective behaviour is to be expected. In Fig. (4), we show a sample of mean square displacement found with this setup. All phases µ$_{mn}$ were conventionally set to zero. Field lines starting locations were picked up from a small interval centered at $\pi/2 \pm \pi/16$. We may fairly well distinguish two regimes: first, the lines are all in phases between them and a collective behaviour follows. This regime is followed up to $L = 10$ (in the units used in the figure) and is characterized by a strong intermittency: the system stays on an attractor, then suddenly jumps on another and so on. Qualitatively, this behaviour may be explained in terms of the kicked oscillator system: while far from the locked modes' location, the bunch of "particles" behaves as a free system following an inertial trajectory (one and the same for all of the particles). When it reaches the location of the locking a sudden perturbation is imparted. As long as the test particles are close enough to each other, the effect of the perturbation is approximately the same over all of them, hence the whole bunch moves rigidly to another trajectory. Of course, over the long run, and quite abruptly, trajectories begin to decorrelate between them, and the usual independent-trajectory behaviour is recovered. The fact that the small-time stair-like trend is a consequence of initial correlations between lines may be seen from Figure (4c). Here, everything is like in Fig. (4b), but for the initial phase of field lines, that are now random. It is apparent how the same behaviour as of Fig. (1) is recovered.

It would be interesting to be able to describe these latter results within the same framework developed in the previous paragraphs. Of course, giving all the dynamics within a single compact expression looks like a tough matter. Hence, we will attempt to describe only the initial development of the system, the stairs-like one (time below approximately 10 units in Fig. 4).

In this case, it appears more convenient to start from Eq. (2): by multiplying both sides by $x^2$ and integrating over $x$, we get

$$\frac{d}{dt}<x^2> = \int_0^t \Lambda(t-\varsigma)\mathcal{L}(\varsigma)d\varsigma \quad , \quad \mathcal{L}(\varsigma) = \int dx\, x^2 \frac{d^2}{dx^2} n(x,\varsigma) \qquad (17)$$

In Eq. (17) the choice $x = 0$ as the starting point is made.

From Fig. (4b) it is apparent that the time derivative in the l.h.s. in Eq. (17) must be zero but for a tiny fraction of time. $\mathcal{L}(\tau)$, on the contrary, is generally different from zero. Hence, $\Lambda$ also must be null almost everywhere. However, we already know that the choice of picking $\Lambda$ as a linear combination of Dirac deltas-which appears rather intuitive-must be ruled out, since it yields the standard diffusive behaviour. The functional form for Λ is therefore rather difficult to devise *a priori* and we must resort to guess it from the data of Fig. (4).

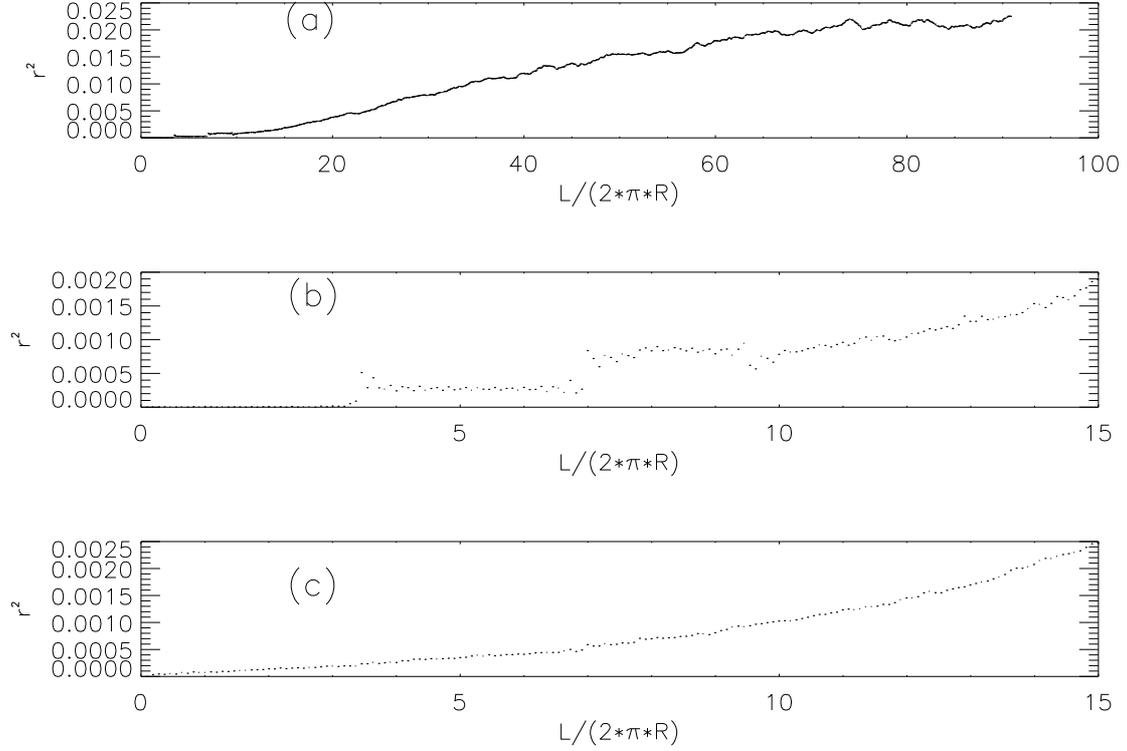

Figure 4. (a) Mean square displacement for a set of trajectories starting at about the same location ($\pi/2 \pm \pi/16$), when the modes are all locked in phase at $\phi = 0$. (b) Zoom at small times. In both plots ε = 0.01 and all other parameters are like in Fig. (1). (c) The same as plot (b), but now the initial locations of trajectories are completely random along ϕ.

We take the Laplace transform of Eq. (17) and get

$$s\tilde{x}^2(s) = \tilde{\Lambda}(s)\tilde{\mathcal{L}}(s) \qquad (18)$$

where $\tilde{x}^2(s)$ is the Laplace transform of $<x^2>(t)$ and we used the fact that, by construction, $<x^2>(t=0) = 0$. It is easy to recognize from Fig. (4) that a reasonable approximation for $<x^2>(t)$ is

$$<x^2>(t) = x_0^2 \sum_m \Theta(t - m\tau) \qquad (19)$$

with Θ the Heaviside function, τ the time between jumps and $x_0$ the magnitude of each jump. We get empirically from Fig. (4) that τ and $x_0$ are approximately constant for all jumps. Because of the definition of $<x^2>$, it is clear that the Θ-dependence must be inherited from $n(x,t)$. Hence, a similar expression must hold also for $\mathcal{L}(t)$, that is, it may be written as:

$$\mathcal{L}(t) = l_0 + l_1 \sum_m \Theta(t - m\tau) \tag{20}$$

The Laplace transforms of (19,20) are

$$\tilde{x}^2(s) = x_0^2 \frac{1}{s} \sum_m \exp(-ms\tau) \approx \frac{x_0^2}{s} \frac{1}{1-\exp(-s\tau)}$$

$$\tilde{\mathcal{L}}(s) = \frac{l_0}{s} + \frac{l_1}{s} \sum_m \exp(-ms\tau) \approx \frac{l_0}{s} + \frac{l_1}{s} \frac{1}{1-\exp(-s\tau)}$$

(21)

In (21) we assume that the sum over *m* consists of sufficiently many terms to make valid taking the limit to infinity. When replaced in (18), expressions (21) yield

$$\tilde{\Lambda}(s) = s \frac{\tilde{x}^2}{\tilde{\mathcal{L}}} = x_0^2 \frac{s}{l_0 + l_1 - l_0 \exp(-s\tau)}$$

$$= \frac{x_0^2}{l_0 + l_1} s \frac{1}{1 - \exp(-s\tau - \xi)}$$

$$= \frac{x_0^2}{l_0 + l_1} \sum_m \exp(-m\xi) s \exp(-ms\tau) \quad , \quad \xi = -\ln\left(\frac{l_0}{l_0 + l_1}\right)$$

$$\rightarrow \Lambda(t) = \frac{x_0^2}{l_0 + l_1} \sum_m \left(\frac{l_0}{l_0 + l_1}\right)^m \Theta(t - m\tau)\delta'(t - m\tau)$$

(22)

The appearance of the derivative of the Dirac delta ($\delta$') is rather unusual; however, note that $\Lambda(t)$ in (22) fulfils our requirement of being null almost everywhere. In order to further check the last expression (22), we suppose $l_0 << l_1$ so that we can retain just the first term in the sum over *m*, that we replace hence into Eq. (2), and get

$$\frac{\partial n}{\partial t} \approx \frac{x_0^2}{l_1} \frac{\partial}{\partial t} \frac{\partial^2}{\partial x^2} n \tag{23}$$

which is satisfied almost everywhere since, by construction, time derivatives are null but for a set of points of zero-measure; hence, it is a consistency check about the validity of our solution. We can qualitatively understand $\Lambda$ of Eq. (22) as follows: Eq. (2) is a particular case of the generic integral equation

$$\frac{df(t)}{dt} = \int_0^t d\varsigma \Lambda(t - \varsigma) g(\varsigma) \tag{24}$$

Le us imagine that $g(t)$ is known. Hence, Eq. (24) tells that the time variation of *f* is determined by $g(t)$ through the kernel $\Lambda$. Let us consider, then, the set $\Omega = \{g(\varsigma)\}, 0 \leq \varsigma \leq t$ and ask: which part of $\Omega$ does actually contribute in the integral (24)? When $\Lambda$ is in the form (12), the whole set contributes, with weights varying and depending just from the time parameter. Instead, in the form (22), not only time but also the form of *g* does matter: Λ picks up from Ω just a few selected elements, in this particular case, only these elements such that $dg/d\varsigma \neq 0$.

In this section we have therefore shown, on a physical case that can be found in different applications (i.e., phase locking phenomena happen in many physical systems), once again the powerfulness of a non-local transport model, although, as we have seen, the task of inferring the transformation kernel can be non trivial at all.

**4. Concluding remarks**

In this work we have presented another contribution to the growing body of evidence according to which anomalous transport in fusion devices can be better described in terms of non-diffusive processes. As pointed out in [12], the use of this relatively new formalism allows describing in a compact and unifying picture disparate observations (in general, all processes whose evolution is inconsistent with diffusive spreading). Since we just dealt with *field line* transport, which is a rather different thing from heat and particle transport, we are not in the position of saying anything about these phenomena, but the present work is a first step into that direction.

The exponential class of memory functions have all moments finite; hence, the Central Limit Theorem holds, according to which, over the long run, one is led back to diffusive behaviour (i.e., to the Gaussian approximation –see chapter 12 in [1]). In other terms, if it were possible to let our system evolve indefinitely, we would find results indistinguishable from ordinary diffusion. But, in our case, finite-size effects occur at finite times, erasing further differences due to memory function. It is not clear to us what the use of one or of another memory function is due to, although we may refer to the papers [27,28], where the Lagrangian correlation function was shown to switch from exponential to power-law form depending upon Kubo number *K* of fluctuations: $K \propto$ (autocorrelation time)/(correlation length)$^2$.

We are aware that our work, just like all the others on the same subject, is merely descriptive. The real issue is about the interpretative capabilities of this formalism. In other terms, whether it is possible to relate (hopefully, in a biunivocal fashion) a specific microscopic physical mechanism to a given analytical expression for the memory function, just like the diffusive equation is the counterpart of a random walk

Markovian microscopic process. Another interesting issue is how it would be possible to define simple quantities (like transport coefficients) which could be compared directly with experimental measurements. To the best of our knowledge, these issues have not yet been broadly addressed in literature, although some works attempting to relate tokamak phenomenology with microscopical generalized evolution equations are starting to appear [29-31].